\documentstyle[aps,pre,epsf]{revtex}
\begin{document}

\draft
\title{Damage Spreading in a 2D Ising Model with Swendsen-Wang Dynamics}
\author{Haye Hinrichsen$^{1,2}$, Eytan Domany$^{1}$
        and Dietrich Stauffer$^{3,4}$}
\address{$^1$Department of Physics of Complex Systems,
         Weizmann Institute, Rehovot 76100, Israel}
\address{$^2$ Max-Planck-Institut f\"ur Physik komplexer Systeme,
         N\"othnitzer Stra{\ss}e 38, 01187 Dresden, Germany}
\address{$^3$ School of Physics and Astronomy,
         Raymond and Beverly Sackler Faculty of Exact Sciences, \\
         Tel Aviv University, Ramat Aviv, Tel Aviv 69978, Israel}
\address{$^4$ Institute for Theoretical Physics, Cologne University,
         50923 K\"oln, Germany}
\date{printed: \today}
\maketitle

\begin{abstract}
Damage spreading for 2D Ising cluster dynamics is investigated
numerically by using random numbers in  a way that
conforms with the notion of submitting the two evolving replicas
to the same thermal noise. Two damage spreading
transitions are found; damage does not spread either at low or
high temperatures. We determine the critical exponents at the
high-temperature transition point, which seem consistent with
directed percolation.
\end{abstract}

\pacs{{\bf PACS numbers:} 05.50.q, 05.70.Ln, 64.60.Ak, 64.60.Ht \\
      {\bf Key words:} \hspace{6mm} Damage spreading, cluster
      algorithms}
\vspace{5mm}
Damage spreading~\cite{Kauf,Creutz,Stanley} turned out to be a useful
tool~\cite{Grass95} to investigate the
dynamics of Ising models. Two replicas of the same system, which
initially differ only on a small subset of the lattice sites,
are simulated using the same random numbers
and one observes how this `damage' spreads during
dynamics by a site-by-site comparison of the two replicas.

Research  done~\cite{JanDarc,Mariz} using the Metropolis, 
Glauber and Heat Bath dynamics, flipping locally
one spin at a time, revealed that whether
damage spreads for a particular model 
or not depends on the kind of dynamics
being used. It was recently demonstrated~\cite{hwd97},
that this ambiguous aspect of damage
spreading can be overcome if one considers all possible single-spin-flip
dynamical procedures that are consistent with the physics of a single
replica. The family of dynamical processes that satisfy this
requirement is quite large;  the above mentioned methods constitute
a small subset of this family.
When more general processes in this familiy 
were considered, damage was shown to spread~\cite{hd97},
even for the one-dimensional Ising model, with exponents
that were either in the Directed Percolation~\cite{GrJSP79} (DP)
or the Parity Conserving~\cite{PC}
universality classes.

In this publication we extend further the dynamical procedures for which
damage spreading is defined. Whereas all work mentioned above was 
done for single-spin-flip dynamics, we study here the evolution when a
non-local procedure, the Swendsen-Wang~(SW) 
cluster algorithm~\cite{SW}, is used.
To our knowledge this was done, so far,
only in Ref.~\cite{Stauffer}, which pointed out a
conceptual difficulty in extending to SW
the definition of ``using the same
random numbers on the two replicas''. We present here
one possible way to overcome this difficulty and address the
issue of how damage spreads when  a non-local algorithm is used.
We also estimate the associated critical exponents.

To understand the difficulty mentioned above note that the SW
algorithm consists of two steps in which random numbers are generated,
namely the construction of the clusters of spins (see below)
and the assignment of their new orientation.  
In DS simulations, however,
even when we use the same random numbers to generate
the SW clusters,  we will in general generate {\it different clusters}
in the two replicas. Ref.~\cite{Stauffer} attempted to associate
clusters of one replica with those of the other by the order of
the clusters' appearance and assigned the same random number
to each such pair of clusters.
Even the number of clusters in the two replicas is generally not the
same; identification by order of appearence
may well cause two groups of spins at very
remote location being assigned the same random numbers. Hence the
observation~\cite{Stauffer} that
after many iterations the two replicas became quite uncorrelated can
be attributed to the fact that the two replicas were, in fact,
not submitted to the same thermal noise.
(This problem can be avoided for the Wolff algorithm~\cite{SW} where our
tests gave always spreading of damage above $T_c$.) 

We propose here another way, one which is more in the
spirit of the standard definition of damage spreading, to deal
with the random number problem for Swendsen-Wang damage dynamics.
We focus on how damage spreads for the Ising model and present
numerical results for $L \times L$ square lattices.

Our method works as follows.
The first step of the SW procedure starts from a 
given spin configuration and
generates clusters. We do this by assigning
a random number $0 \leq p_{ij} \leq 1$ to every {\it bond}, 
i.e., the same
number is assigned to a given bond on the two replicas.
Each bond is either frozen or deleted according to the
standard SW rule~\cite{SW}:
\begin{quote}
If $S_iS_j=-1$ the bond is deleted. \\
If $S_iS_j=1$ it is deleted if $p_{ij} \leq {\rm exp}(-2J/k_BT)$
and frozen otherwise. \\
Sites connected by frozen bonds form the SW clusters.
\end{quote}
With this method, 
identical spin configurations on the two replicas clearly give rise to
identical clusters.

The second step of the SW procedure assigns randomly to
each cluster a spin value, $\pm 1$, by assigning a single
random number to each cluster. The problem 
mentioned above arises at this step; the number
of clusters may differ in the two replicas, any spin can belong to one
cluster in one replica and to a different one in the other. 
What is the meaning, in this situation, 
of using the same random number for the two replicas?
To overcome this ambiguity, we assign different 
random numbers  $p_i$ to each site of the lattice with the 
constraint that the same random number is assigned to the same
site $i$ of the two replicas. The $p_i$ are uniformly distributed
about zero and the new status of every cluster
is determined depending on whether the
{\it sum of all the random numbers assigned to its
sites} is positive or negative. Clearly, if two 
clusters in the two replicas contain identical sites, 
these sums will be identical, and the two
clusters will be treated in the same way. 
If the two clusters are nearly identical, 
then the two sums will be strongly correlated and most probably
treated the same way. If the two clusters 
share no common site, the two sums will be completely 
uncorrelated and so will be the cluster orientations. This
completes the description of a single SW step, which is our
Monte Carlo time unit.
New random numbers are chosen after every step\footnote{
It is natural to associate the random numbers 
with thermal noise which, in turn,
is normally assumed to be local in 
space and time. In this sense the
procedure we propose is more "physical" 
than the one used in~\cite{Stauffer}.
}.

The procedure outlined above is the analogue of the  Heat Bath
algorithm, to which we restrict our attention in this 
the following of this paper. After equilibration
we  introduce  damage by flipping in one replica sites that belong to
one line in the center of the lattice. Only the odd 
sites on it are damaged to prevent the `infinite' cluster
from splitting into two halves below the critical temperature $T_c$. 
(We used helical boundaries in one direction and free boundaries in the 
other. We also initially damaged the center quarter
of the whole lattice, and got the same spreading temperature as given
below.)

Fig.~1a shows the equilibrium damage as a function of temperature.
As expected, damage does not spread at low temperatures;
it starts to spread below $T_c$ and  its limiting long-time value
reaches a maximum at $T_c$. For $T \leq T_c$ we find enormous
fluctuations, and even extended regions
of damage may vanish completely within a single time step.
We ascribe this to the presence of an `infinite' cluster below $T_c$,
which we flip just as we do with many finite clusters.
Interestingly, there is a second transition --
damage shrinks and vanishes above a spreading temperature $T_s$,
with $1.33 < T_s/T_c < 1.34$ for $200 \le L \le 1000$.
Close to $T_s$ the damage seems to
vanish as $(T_s-T)^\beta$ with $\beta = 0.65$; see Fig.~1b.

The Swendsen-Wang dynamics was invented to reduce critical slowing down;
indeed, right at $T=T_c$ the relaxation time in two dimensions increases
only logarithmically with system size~\cite{Tamayo}.
Our damage, however, shows strong critical slowing down
on both sides of the spreading temperature $T_s$.
For a quantitative study  we use the damage vanishing
method~\cite{Grass95,Matz}: Initially
half of the lattice is damaged (if all of it is damaged, the damage
vanishes immediately). Next, we checked 
for $T > T_s$ how the damage decays to zero;
Fig.~2a shows that it does so exponentially,
after an initial transient.
Fig.~2b shows that the corresponding 
exponential relaxation times are roughly given by
$\tau_r \propto (T/T_c - 1.335)^{-1.2}$.
When, however, instead of $\tau_r$ we studied $\tau_1$, the average time
after which the damage has become exactly zero (not shown), an exponent
close to 0.9 was observed; such discrepancies have
been observed earlier~\cite{Matz} with the
latter definition of relaxation times.

We also performed more limited studies of damage spreading from a single
initial site. In this type of simulation~\cite{Grass89}
one usually expects the damage to survive
after $t$ iterations with a probability 
proportional to $t^{-\delta}$, the number of damaged sites
to grow as $t^{\eta+\delta}$, and the mean square distance of
the damaged region from the origin of the damage to grow as $t^{z}$.
The latter two quantities are averaged only over those 
lattices which are still damaged at time $t$.

The initialization with a single seed of damage
is, however, very inefficient numerically, 
since often the damage vanishes very fast 
and we end up simulating a whole lattice to study
only a small region around this site. Hence to 
get meaningful results one must average the evolution 
over a very large number of sample runs which, in turn,
limits the sizes of the lattices used. 
For example, we performed averages
over 5000 runs of an $301 \times 301$ system; 
the results obtained this way may, therefore, 
be strongly influenced by finite size effects. With all these
caveats taken into account we obtained (see Fig.~3) 
at $T = T_s$ the following results:
{\it a)} The survival probability of damage to time $t$ is  
$\propto t^{-0.5}$,
i.e. we get   $\delta=0.5$ (versus the DP 
result~\cite{Grass89} $\delta=0.460(6)$); {\it b)} the number of damaged
sites  grows as $t^{0.7}$ (to be compared with $\delta + \eta =0.681$ in DP -
note that we measure damage per surviving runs); 
{\it c)} The mean square distance of
the damaged region from the origin of the damage grows as $t^{1.1_5}$,
whereas $z=1.134$ in DP. Since our exponents 
(including our $\beta =0.65$, vs 0.584) deviate from those of
DP~\cite{Grass89} by only about $10 \%$, we believe that  
the observed damage spreading transition at $T=T_s$
belongs to the universality class of directed percolation.


In summary, we studied damage spreading using a non-local algorithm.
By introducing a definition of damage spreading for the Swendsen-Wang
algorithm which conforms with the standard notion of submitting the two
replicas to the same thermal noise, in contrast with
Ref.~\cite{Stauffer}, we discovered some
non-trivial results; damage spreads at temperatures {\it between} two
transitions. Since at $T_s > T_c$
the clusters are finite the transition 
according to our numerical estimates of
the exponents may well be in the standard
DP universality class; below $T_c$, however, 
`infinite' clusters are present
and the damage spreading transition 
may be in a new universality class.
Possible extensions of this
work include working at higher dimensions, with
different spin models, as well as improving the accuracy
of our numerics on the square lattice.

We thank the German Israeli Foundation for partial support;
DS thanks Joan Adler for hospitality at the Technion in Haifa, where
part of the work was done.

\centerline{Figure Captions}

\vspace{2mm}
Fig.1: Fraction of damaged sites in equilibrium versus temperature. 
Graph a shows the whole investigated region on linear scales, 
graph b persents a double-logarithmically plot of the same data, 
slightly below the spreading temperature.

\vspace{2mm}
Fig.2: Relaxation of the damage for $1.35 \le T/T_c \le 1.60$, 
i.e., above the spreading temperature, 
depending on the distance from $T_s$, damage is
introduced at time = 30, 50, or 100. Straight lines in the
semilogarithmic plots of graph a correspond to exp$(-t/\tau)$, 
and the relaxation times $\tau$ are
plotted double-logarithmically in graph b.

\vspace{2mm}
Fig.3: Dynamics at the spreading temperature $T/T_c \simeq 1.335$: 
Number of surviving damages ($\diamond$) out of 5000 samples, 
mean square distance (+) and number of damaged sites ($\Box$). 
The latter two quantities are summed up
over all samples and then divided by the 
number of surviving damages ($\diamond$),
i.e., by the number samples which were 
still damaged at time $t$.


\begin{figure}
\epsfxsize=155mm     
\centerline{ \epsffile{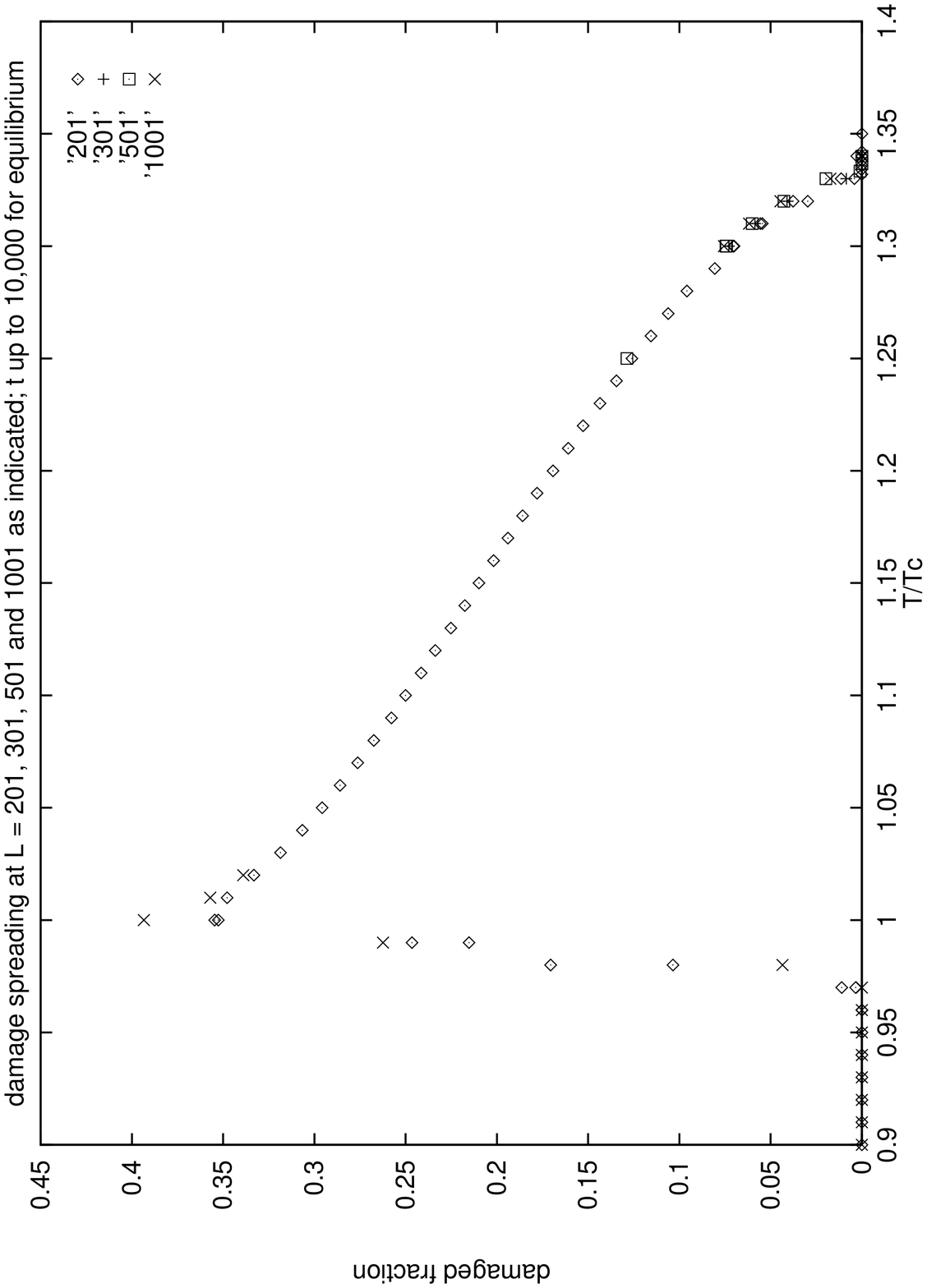}  } 
\centerline{ Figure 1a }
\end{figure}

\begin{figure}
\epsfxsize=155mm     
\centerline{ \epsffile{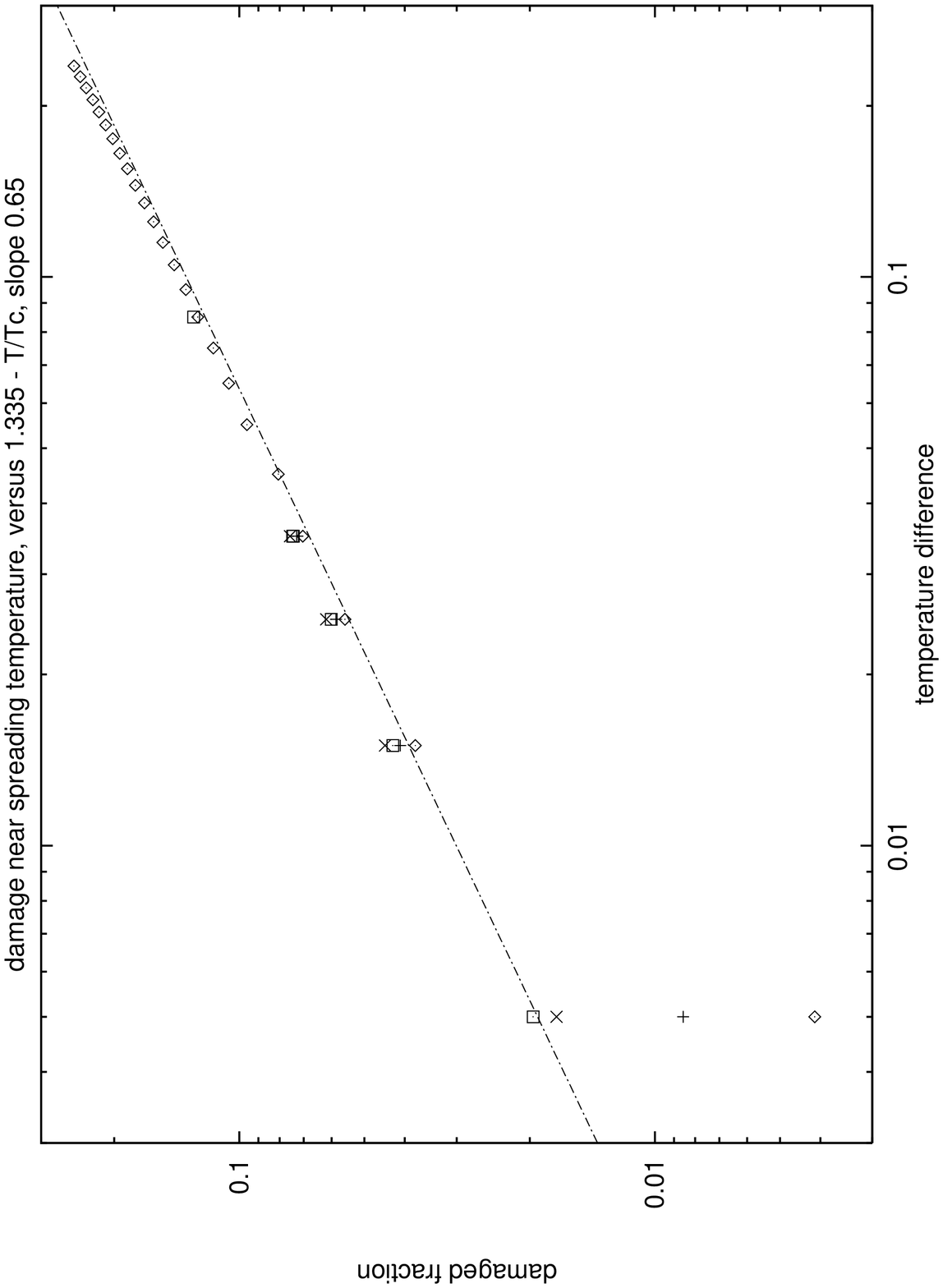}  } 
\centerline{ Figure 1b }
\end{figure}

\begin{figure}
\epsfxsize=155mm     
\centerline{ \epsffile{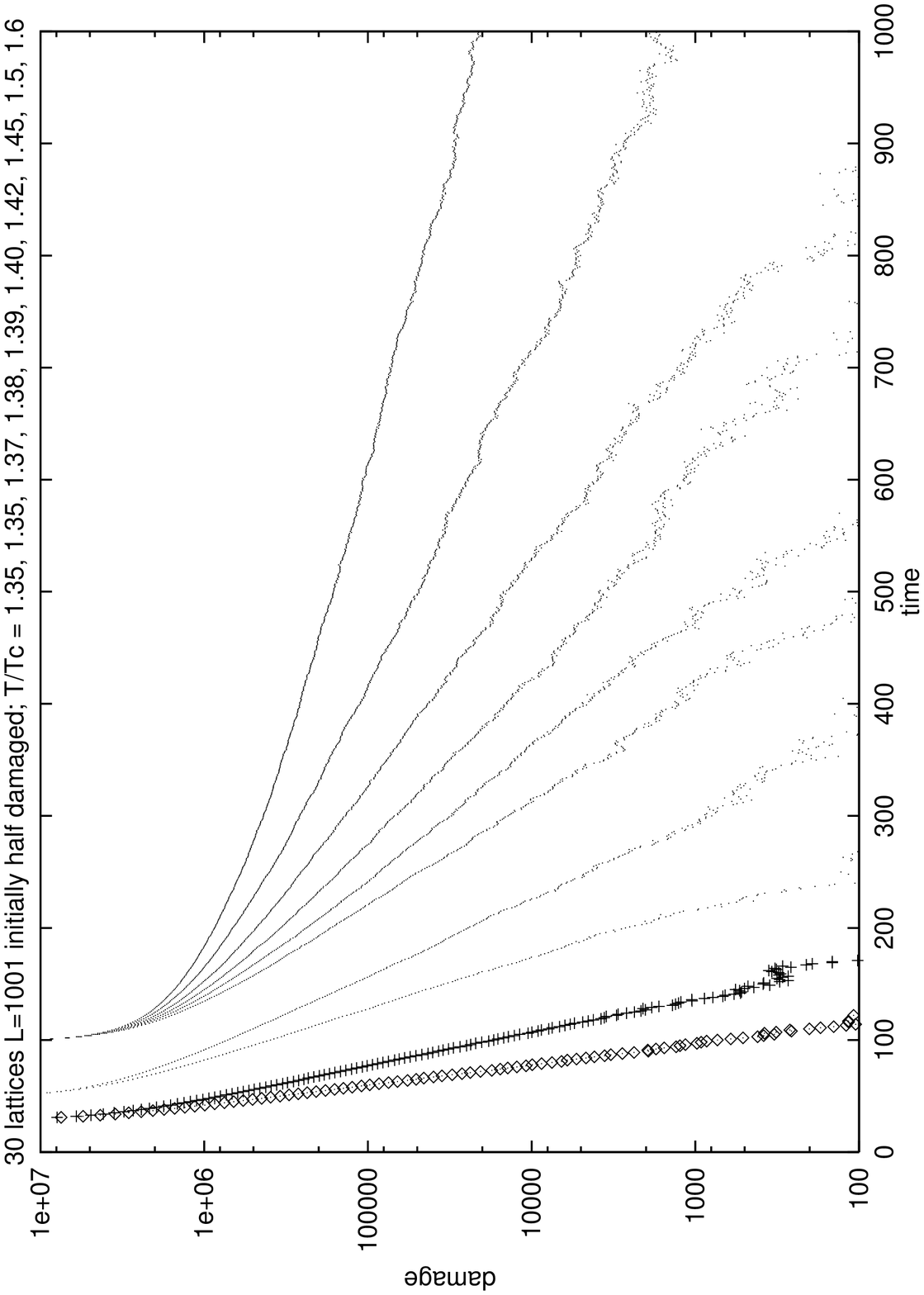}  } 
\centerline{ Figure 2a }
\end{figure}

\begin{figure}
\epsfxsize=155mm     
\centerline{ \epsffile{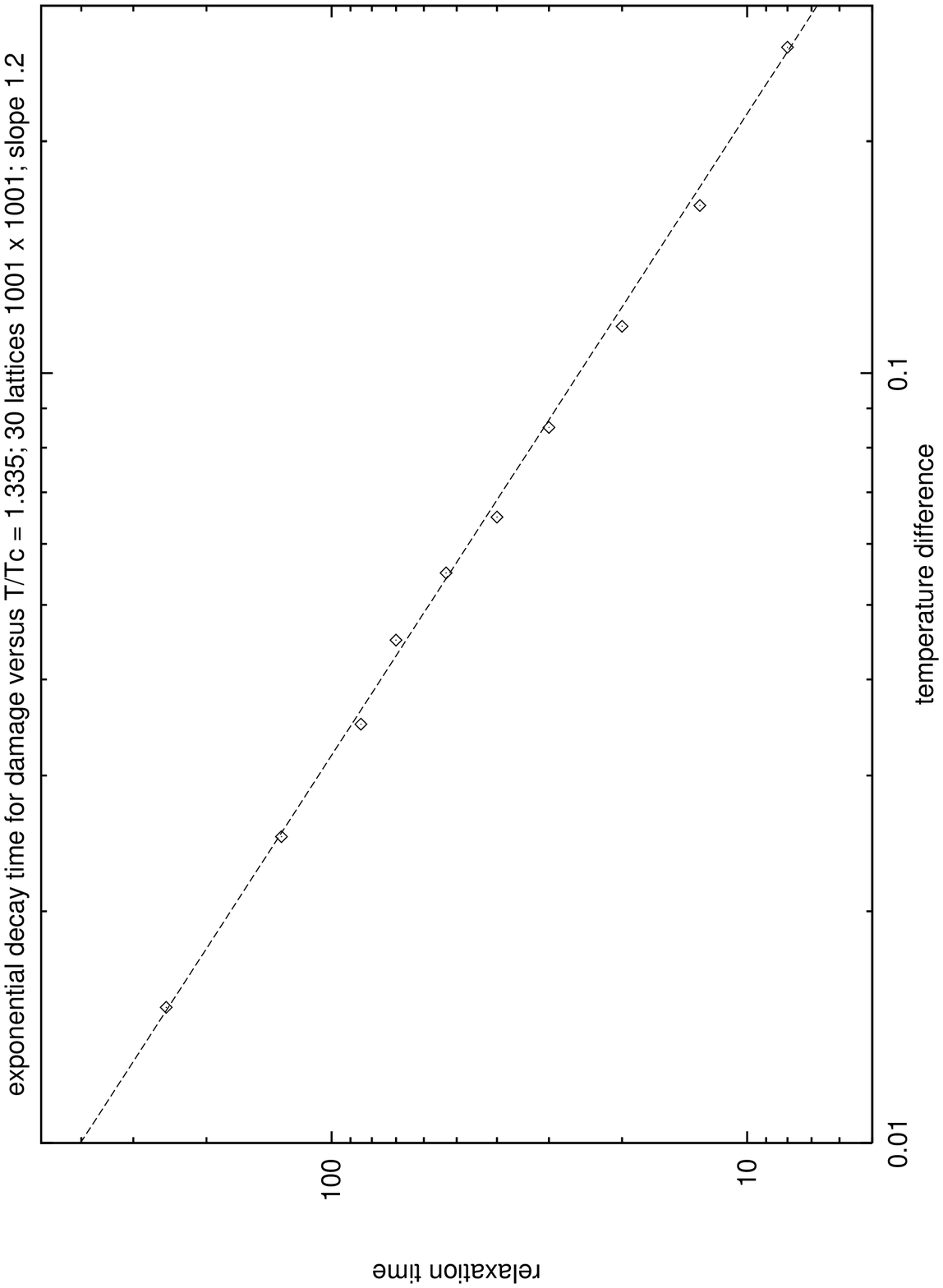}  } 
\centerline{ Figure 2b }
\end{figure}

\begin{figure}
\epsfxsize=155mm     
\centerline{ \epsffile{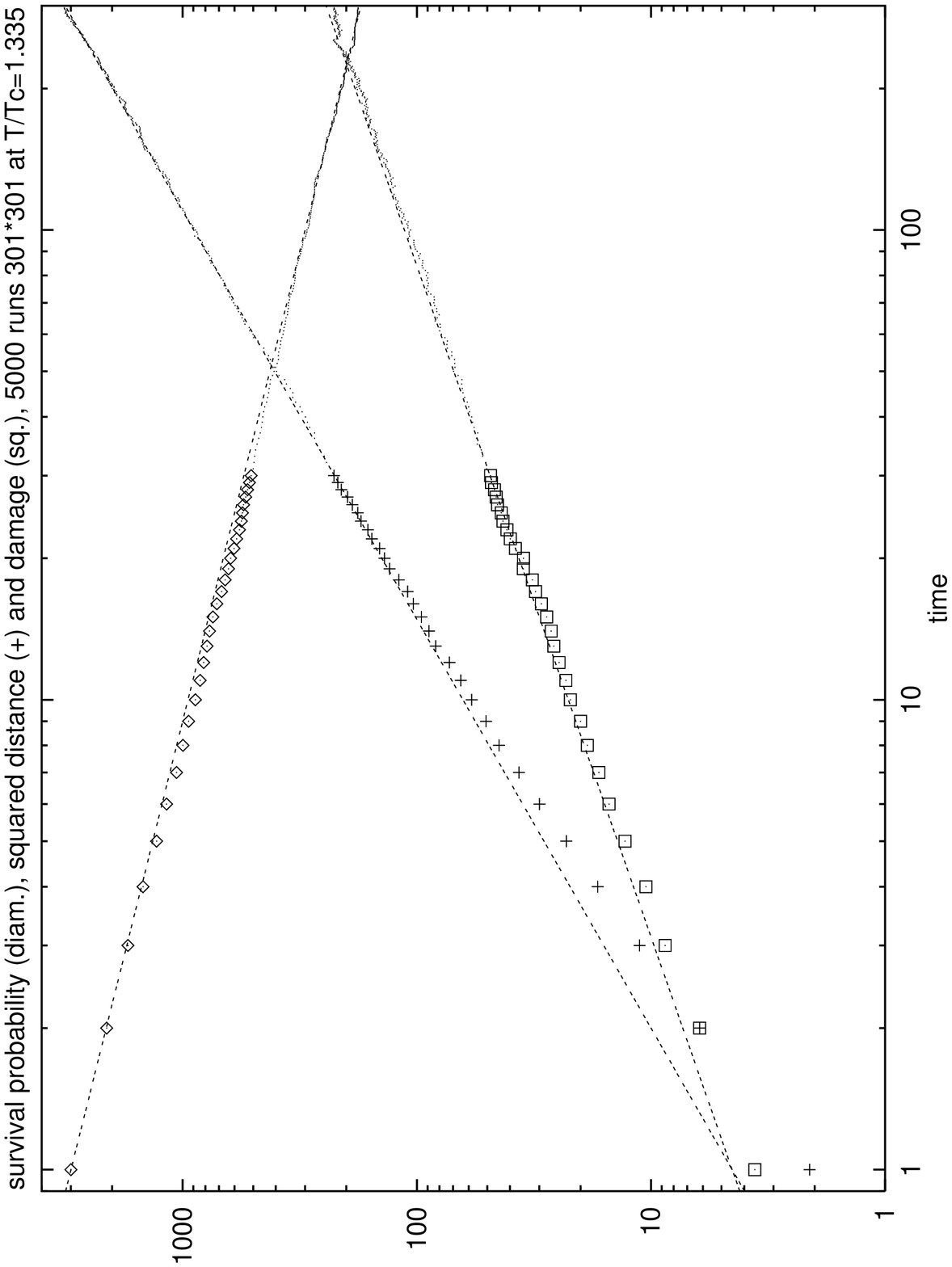}  } 
\centerline{ Figure 3 }
\end{figure}

\end{document}